\documentclass[mathleft]{an}
\usepackage{graphicx,times}
\usepackage{bm}
\usepackage{rotating}
\newcommand{\EQ}{\begin{equation}}
\newcommand{\EN}{\end{equation}}
\newcommand{\EQA}{\begin{eqnarray}}
\newcommand{\ENA}{\end{eqnarray}}

{}
{}
{}

{}
{}
{}
{}
{}
{}
{}
{}
{}
{}
{}
{}
{}
{}
{}
{}
{}

{}
{}
{}

{}

{}
{}

\Pagespan{797}{}
\Yearpublication{2009}
\Yearsubmission{2009}
\Month{7}
\Volume{330}
\Issue{8}
\DOI{10.1002/asna.200911242}
\title{The meaning inferred from the spin period distribution of normal pulsars}
\author{Ying-Chun Wei
          \inst{1}
          \and
          Cheng-Min Zhang
          \inst{1}
          \and
          Yong-Heng Zhao
          \inst{1}
          \and
          Qiu-He Peng
          \inst{2}
          \and
          Xin-Ji Wu
          \inst{3,}
          \inst{4}
          \and
          A-Li Luo
          \inst{1}
          \fnmsep\thanks{Corresponding
          author: zhangcm@bao.ac.cn;
ycwei@bao.ac.cn}} \institute{ National Astronomical Observatories,
Chinese Academy of Sciences, Beijing 100012, China \and Department
of Astronomy, Nanjing University, Nanjing 210093, China \and
National Astronomical Observatories/Urumqi Observatory, Chinese
Academy of Sciences, Urumqi 830011, China \and Department of
Astronomy, Peking University, Beijing 100871, China } \received{2009
} \accepted{ } \publonline{ }  \abstract{ We make statistics about the number
distribution of normal radio pulsars according to the pulsar spin
period. And find the physical meaning of the traditional statistical
method about this problem is very limited. According to the
statistical method proposed by us, the distribution rule is exactly
the exponentially decay with the
 spin period increasing, especially important, which can be
interpreted as normal radio pulsar number decays as the radiative
elements decay. It discloses that though belonging to two different
extreme scales of the universe and having different inner active
mechanisms, the normal radio pulsars and radiative elements accord
with the same varying rule. And like the half decay period of the
radiative elements we can also introduce the half decay spin period
of the normal radio pulsars. }

\begin{document}
\maketitle

\section{Introduction}

Pulsars---a kind of charming compact objects in the sky, are
generally known to be born in the supernova exploration. Exploring
them will provide the opportunity for people to understand the
fundamental physics and astrophysics (Cordes 2004). There are many
questions about pulsars deserved to be studied. In this paper we
would concentrate on one of the attractive problems that is how the
pulsar number distributes according to the spin period. In section 2
we introduce the usual statistical method at present on the spin
periods of normal pulsars; in section 3 we point out the problem
existing in this present statistical method, and introduce the
concept of ``generation" of normal pulsars; in section 4 we
introduce a new statistical method on the spin period distribution of
normal pulsars; in section 5 we make a conclusion.

\section{Usual Statistical Method on the Spin Periods of Normal Pulsars at Present}

\begin{figure}

\centerline{\includegraphics[angle=0,width=8cm]{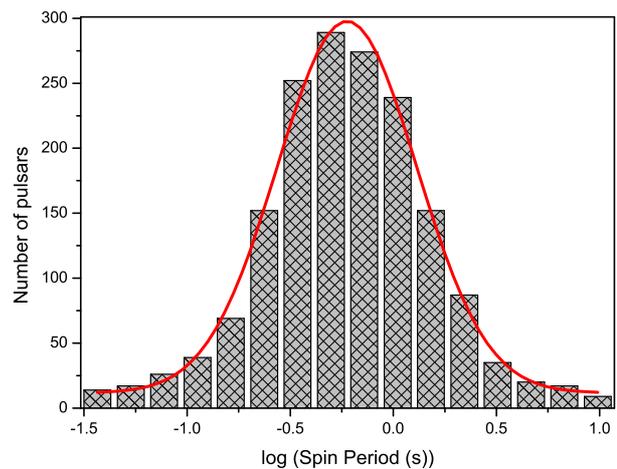}}
\caption{The normal pulsar number distribution according to the log
values of spin periods. The data are from the ATNF pulsar catalogue
for pulsars with spin periods larger than 30 ms. See the internet
Web: http://www.atnf.csiro.au/research/pulsars/psrcat (Manchester et
al. 2005).} \label{fig:taud}
\end{figure}
As is well known pulsar spin periods usually span three magnitudes
or so from milliseconds to seconds. So people like to do statistics
for the log values of the spin periods (Manchester, Hobbs, Teoh \&
Hobbs 2005; Manchester 2009), or directly under the log coordinate
of the spin periods (Manchester 2009), because the log function can
transform the different scales of spin periods to the same scale. In
Fig.~\ref{fig:taud} we illustrate this method for the normal
pulsars whose spin periods exceed 30 ms (Lyne \& Smith 2006) in the
ATNF pulsar catalogue (Manchester et al. 2005). It seems to accord
with the following Gaussian distribution very well:
\begin{equation}
y=y0 + \frac{A}{w\sqrt{\pi/2}}e^{-2(\frac{x-xc}{w})^2},
\end{equation}
where $y0=11.741\pm3.963$, $A=243.163\pm8.562$, $w=0.678\pm0.020$,
$xc=-0.225\pm0.008$ are the fitting parameters. The coefficient of
determination (COD) is 0.99438. The smaller the relative errors and
the closer to 1 the COD, the better the fitting. Apart from $y0$,
the relative errors of the other parameters are all less than $5\%$.
So it seems safely to say that the normal pulsars distribute
Gaussianly according to the log values of spin periods.

\section{Concept of ``Generations" of Pulsars}

But we should not forget that these normal pulsars were not born in
the same period. In fact, they are born with some birthrate $1\sim3$
per century (Vranesevic et al. 2004; Faucher-Gigu\`{e}re \& Kaspi
2006) which can be larger five times taking into the uncertainties
to form the current of pulsars in the Galaxy. At the beginning of
the current, i.e., the newborn pulsars have a distribution of the
spin periods which nowadays is regarded to be about 0.1-0.5 s for up
to $40\%$ of all pulsars (Vranesevic et al. 2004; Lorimer 2006). So
maybe we could guess that the distribution of spin periods of
newborn normal pulsars is pulse form like Gaussian distribution with
the distribution peak at 0.1-0.5 s. Certainly, the distribution
pulse form of spin periods of the newborn pulsars could not approach
to the 0 infinitely because there is a smallest rotation period of
pulsars smaller than which the pulsar will be broken up and which
can approximately be estimated as small as 1.5 ms when the
centrifugal force is just equal to the gravity at the equator (Lyne
\& Smith 2006).

\begin{figure}

\centerline{\includegraphics[angle=0,width=8cm]{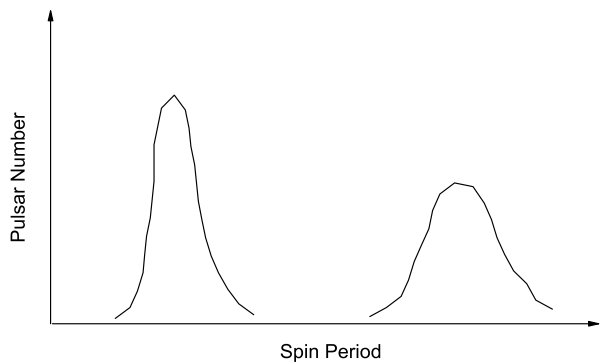}}
\caption{The sketch map of different pulsar generations at different
spin periods.} \label{fig:taud2}
\end{figure}

   In the initial part of the current of pulsars which can be imaged as
``initial generation of pulsars", the spin period distribution
should be like the spin period distribution of newborn pulsars. Due
to the radiative energy taking out the rotation energy of pulsars
(Gold 1968, 1969; Pacini 1968), the rotation of pulsars will turn
slow gradually and this pulse-like distribution of spin periods of
the initial generation will move in the right direction of spin
period axis, certainly with the possibility that some pulsars could
not radiate the radio pulses owing to the extinction of the polar
cascade process (Chen \& Ruderman 1993; Hibschman, Johann \& Arons
2001) and the pulse distribution form may also change as illustrated in
Fig.~\ref{fig:taud2}.

   Let us focus on the statistical step sizes in Fig.~\ref{fig:taud}. The peak in
Fig.~\ref{fig:taud} is at $10^{-0.225}\approx 0.6$ s approaching to the
assumed 0.1-0.5 s magnitude of the distribution peak of the spin
periods of the initial generation pulsars. So we can guess there are
not many generations of pulsars left to the peak in
Fig.~\ref{fig:taud}. In addition the statistical step size left to
the peak is even so smaller that it could not contain one pulsar
generation and in fact to some extent reflect on the distribution
pulse form inner structure of spin periods of the initial several
generations of pulsars. Whereas the statistical step sizes right to
the peak in Fig.~\ref{fig:taud} are generally very large that can
contain a few generations in the same statistical step. So in fact
the statistics right to the peak in Fig.~\ref{fig:taud} reflects
the change rule of pulsar numbers of different generations with spin
periods.

Now we make clear that the statistical step sizes in
Fig.~\ref{fig:taud} are suitable to do different things that reflect
 different physical meanings: one is for the distribution of spin
periods in the pulse-distribution-form of the initial generations; the other is for the distribution
of pulsar numbers among different pulsar generations. The log
transformation mixes the different study scales of the two different
questions, consequently confuses the two questions. The result is
that the physical meaning of Gaussian distribution in
Fig.~\ref{fig:taud} is very limited.

\section{New Statistical Method on the Spin Periods of Normal Pulsars}

\begin{figure}

\centerline{\includegraphics[angle=0,width=8cm]{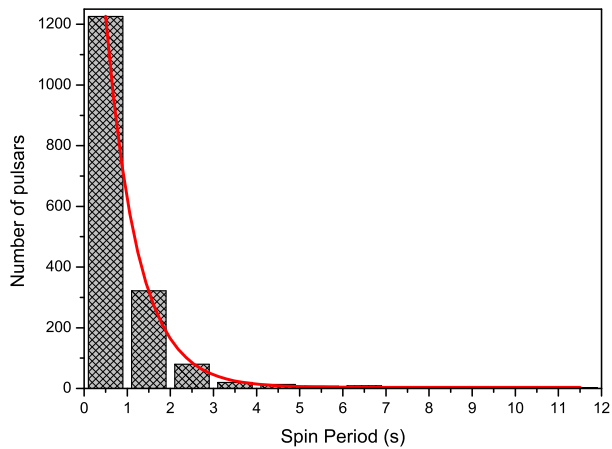}}
\caption{The pulsar number distribution in different pulsar
generations. The samples are the same as in Fig.~\ref{fig:taud}.}
\label{fig:taud3}
\end{figure}

Since nowadays the pulsars observed are including various
generations, we can study the pulsar number distribution among
different pulsar generations, i.e., we would study the pulsar number
distribution with the spin periods, not with the log values of spin
periods. Fig.~\ref{fig:taud3} is the result. It accord with the
following exponential decay with the very surprising COD 0.99992:
\begin{equation}
y = A1~e^{-x/t1} + c,
\end{equation}
where $A1=2408.115\pm15.750$, $t1=0.738\pm0.006$ s,
$c=3.614\pm1.203$ in Fig.~\ref{fig:taud3}. The fitting of
Fig.~\ref{fig:taud3} is obviously better than that of
Fig.~\ref{fig:taud}.
We know for pulsars the larger the age, the longer the spin period,
and the weaker the radiative ability. Fig.~\ref{fig:taud3} tells us
that the active pulsar number would gradually reduce with the
increasing of spin periods. Meanwhile it also tells the active
pulsar number drops with the increasing of time. If the spin period
has linear relationship with the age, then the active pulsar number
would also drop exponentially with time. But the spin periods do not
have simple relationship with the time (Lyne \& Smith 2006;
Camenzind 2007; Gonthier, Van Guilder \& Harding 2004; Contopoulos
\& Spitkovsky 2006). In spite of this we still can introduce the
half-decay spin period for pulsars like the half-decay period for
radiative elements: the value from Fig.~\ref{fig:taud3} is about
$0.516\pm0.004$ s.

\section{Conclusion}

The pulsars and radiative elements belong to very different scales
of the universe respectively, and have very different inner active
mechanisms, whereas they have such similar decay rules, i.e., their
activities drop exponentially though with different variables: one
with the spin period; the other with the time.

Fig.~\ref{fig:taud3} does not take into count the selective effect
of observation, the most important selective effect is the
luminosity, because the radiative capability of pulsars declines
with age. It is very difficult to count this selective effect.
Besides this, we do not forget that pulsars are the high velocity
objects (Vranesevic et al. 2004; Arzoumanian, Chernoff \& Cordes
2002; Hobbs, Lorimer, Lyne \& Kramer 2005), which would make the
number of pulsars in some region of the sky have some change. So
Taking into account various effects will bring very large
uncertainties. In spite of all these, we still believe that the law
disclosed in Fig.~\ref{fig:taud3} is true, partly from the
philosophical consideration that the different scales in the
universe might abide by some similar law.

\acknowledgements We thank G.J. Qiao, K.F. Wu, X.Y. Chen, H.B.
Zhang, L.D. Zhang and J.J. Zhou for helpful discussions. This
research has been supported by NSF of China (No.10778611,
No.10773017, No.10973021 and No. 10573026) and National Basic
Research Program of China (No. 2009CB824800). The authors express
the sincere thanks to the critical comments.

\end{document}